# The Fizeau Interferometer Testbed

Xiaolei Zhang
US Naval Research Laboratory, Remote Sensing Division
Washington, DC 20375, USA
Tel: (202) 404-2389, email: xiaolei.zhang@nrl.navy.mil

Kenneth G. Carpenter, Richard G. Lyon
NASA's GSFC, Greenbelt, MD 20771, USA

Hubert Huet, Joe Marzouk
Sigma Space Corporation
9801 Greenbelt Rd, Suite 105, Lanham, MD 20771, USA

Gregory Solyar
GEST/UMBC, NASA's GSFC, Greenbelt, MD 20771, USA

*Abstract*—The Fizeau Interferometer Testbed (FIT) is a collaborative effort between NASA's Goddard Space Flight Center, the Naval Research Laboratory, Sigma Space Corporation, and the University of Maryland. The testbed will be used to explore the principles of and the requirements for the full, as well as the pathfinder, Stellar Imager mission concept. It has a long term goal of demonstrating closed-loop control of a sparse array of numerous articulated mirrors to keep optical beams in phase and optimize interferometric synthesis imaging. In this paper we present the optical and data acquisition system design of the testbed, and discuss the wavefront sensing and control algorithms to be used. Currently we have completed the initial design and hardware procurement for the FIT. The assembly and testing of the Testbed will be underway at Goddard's Instrument Development Lab in the coming months.

TABLE OF CONTENTS

## 1. INTRODUCTION

The Fizeau Interferometer Testbed is a ground-based testbed for Stellar Imager (SI, http://hires.gsfc.nasa.gov/∼si) [1], [2], an UV-optical interferometry mission in NASA's Sun-Earth Connection far-horizon roadmap currently undergoing concept and design study at NASA's Goddard Space Flight Center. The primary science goals of the SI include the studying of the spatial and temporal stellar magnetic activity pattern in a sample of stars, which will enable improved forecasting of solar activity as well as improved understanding of the impact of stellar magnetic activity on planetary climate and astrobiology; the measurement of internal stellar structure and rotation using the technique of asteroseismology and the determination of their relation to the stellar dynamos; as well as the imaging of the stars in external planetary systems which allows the determination of the impact of stellar activity on the habitability of the surrounding planets, thus complementing the assessment of external solar systems that will be done by the planet finding and imaging missions, such as SIM, TPF and PI.

The mission calls for a reconfigurable array of 10-30 one-meter class spherical mirrors to be used in a Fizeau, or image plane beam combination mode. The maximum baseline length is about 500 meters. The wavelength range of operation will be in the optical for asteroseismology and in the ultraviolet for surface imaging, including two of the important emission lines for studying the stellar dynamo behavior, i.e. the chromospheric Mg II h&k lines near 2800 Å and the transition region C IV doublet at 1550 Å. The best angular resolution achievable is about 60 micro-arcsec at 1550 Å, which corresponds to about 40,000 km linear resolution for a sun-like star at a distance of 4 parsec. There will be roughly 30 linear resolution elements at the equator of a typical nearby dwarf star, and about 1000 resolution el-ements to cover the stellar surface. The spectroscopic capability includes passbands as narrow as a few Angstroms up to hundreds of Angstroms, from the UV up into optical wavelengths. It is intended to be a long-term (∼10 year) mission for the studying of the stellar magnetic activity cycles.

In what follows we first give an example of the sensitivity calculation for a typical SI target star.

The signal-to-noise ratio $R$ for photon-counting detector is

$$R = (\bar{n}T)^{1/2} \tag{1}$$

where $\bar{n}$ is the average number of received photons per unit time, and T is the integration time.

A typical target star for SI is Procyon, which has a distance of 3.48 pc from the Sun, and a linear size of $2.38 \times 10^6$ km. For the C IV line at 1150 A, the frequency of the line is $\nu = c/\lambda = 3 \times 10^{15}$ Hz, and the energy of the photon at this frequency is $E = h\nu = 2 \times 10^{-11}$ ergs. The measured energy flux at the C IV line is F= $111 \times 10^{-13}$ ergs/cm²/sec. For a single 1 meter diameter collector mirror, the flux of the number of photons collected by the primary beam (ignore losses) is thus

$$\bar{n}_1 = \frac{F * A}{E} = 4375 \text{ photons/sec.} \tag{2}$$

For N collectors, the total received flux = $N\bar{n}_1$. For unresolved point sources, all of the received flux by the N collectors, in the Fizeau beam combination mode, falls onto a single pixel. If the object is resolved into M pixels, however, the flux per pixel is $\bar{n}_1 N/M$.

For Procyon, with its distance $d = 3.48$ pc and diameter $D = 2.38 \times 10^6$ km, its angular diameter is $D/d = 2 \times 10^{-8}$ radian. The angular size of the SI primary beam, for 1 meter collector mirror, is $1.22\lambda/1(m) = 1.2 \times 10^{-7}$ radian. Therefore the primary beam is about 6 times bigger in linear dimension than the steller diameter of Procyon. The total linear resolution element within the single-dish Airy disk, assuming a 500 meter baseline and 1 meter mirrors, is 500, out of which 1/6 of it falls onto the stellar surface. The total number of areal resolution elements covering the stellar surface is thus $(500/6)^2 = 6944$. The flux per resolved pixel received by a 30-element interferometer is thus

$$\frac{\bar{n}_1 \times \text{efficiency}}{6944} = 2.3 \text{ photons/sec/pixel} \tag{3}$$

where we have assumed a 10% overall telescope efficiency. For S/N = 5, the integration time required is about 10 seconds for photon-noise dominated detection. This is a reasonable amount of integration time for science observations.

For co-phasing of the array, if an image-based wavefront sensing approach is to be used, continuum observation in the optical band can provide up to several orders of magnitude improvement in sensitivity over that of the line observation. The achievement of a high-bandwidth control loop, however, will require in addition a fast frame rate detector array as well as fast numerical algorithms to process the imaging data and derive the updating control parameters. Thorough evaluations of the relative merits of wavefront sensing versus direct metrology approaches in maintaining the alignment accuracy of the interferometer array will need to be carried out.

It is clear that the realization of long-term goals of the Stellar Imager mission requires unique architectural as well as technological explorations. The Fizeau Interferometer Testbed will serve as an important first step towards reaching the goals and requirements of the Stellar Imager.

## 2. FIZEAU VERSUS MICHELSON BEAM COMBINATION

During the initial conception phase of SI, both the Michelson (pupil plane beam combination) and the Fizeau (image plane beam combination) modes have been considered. The Fizeau configuration was chosen in the end because of several advantages it offered for a mission like SI. Since the Michelson approach requires that the beams from all of the elements be combined and interfered pairwise with each of the other beams, the total number of elements is limited to 10 or less in order to avoid overly complicated beam combiner designs. The Michelson option thus requires numerous reconfigurations of the array to obtain a full baseline coverage. The Fizeau approach, on the other hand, could possibly utilize a much larger number (∼ 30) of simpler and less expensive one-meter class flat or spherical mirrors on microsats, distributed on a spherical or paraboloidal surface. The light beams from all the elements would be combined simultaneously on one detector; alternatively they could be picked up and combined in subsets if desired. This option requires far fewer reconfigurations (i.e. 2 instead of 20) to obtain a synthesized image, which should save both time and propellant. This option should also utilize fewer reflections, an important consideration if the facility is to operate in the ultraviolet.

Despite the recent surge of development effort in both ground-based as well as space-based interferometry, motivated in part by the needs of the various planet finding and imaging missions, most of the effort has so far been focused on the development of the Michelson type of interferometers, with comparatively little effort for the Fizeau type. This disparity is partly due to the myth that "if Michelson interferometry is hard, Fizeau interferometry is *impossibly* hard". It is true that while Michelson interferometry in general requires only the *knowledge* of the baselines and optical paths to a fraction of the observing wavelength, Fizeau interferometry generally requires the *control* of these same parameters to a fraction of a wavelength, at least in the direct imaging mode. For ground-based applications, this increased accuracy requirement on the control of pathlengths and baselines also translates to a fast and high-accuracy sensing of the wavefront (or optical alignment) in order to derive high bandwidth control signals to combat the fast-changing atmosphere.

In the space environment, on the other hand, one no longer has to worry about the atmosphere-induced wavefront changes, the wavefront variations will result mostly from the vibrations due to the station-keeping mechanisms as well as thermal effects. These effects are more predictable in nature, and laser metrology of parts of the system susceptible to mechanical vibrations can be employed to aid the overall wavefront sensing effort. In the end, how much of the wavefront sensing will be done by image-based methods and how much by direct metrology will have to be determined empirically by the effectiveness and efficiency of each approach, as well as by the technological evolution of the vibration isolation methods.

We would like to comment here that once again the Fizeau configuration here offer the flexibility of using any subset of the full array mirrors to do partial beam combination without the need to drastically change the hardware configuration. In one extreme, if one were to adopt pair-wise beam combination, the 30 mirror SI array will involve about 435 separate baseline observations. However, given the large mirror size (1 m) and the superior sensitivity it offers (10 seconds average integration time per baseline for science observations), the 435 baselines can be observed in 1.2 hours integration time. Even after taking into account the overhead of reconfiguration and cophasing of the array after a source change, one can still finish observing an average star within a fraction of a day. Considering the 10 year mission lifetime and the about 1000 or so target stars, even a pair-wise observation mode can achieve most of the mission objectives well within the mission lifetime. To phase up two mirrors at a time is of course well within the state of the art of the current space interferometry technology.

There is, however, one class of the observation which will be hindered by this slower mode of operation, i.e., those of rapidly rotating stars. Therefore, one still would like to phase up as many mirrors at one time as possible within a fraction of the science integration time, all the way up to the entire 30 elements, which is still not out of the question given the superb continuum sensitivity of SI and thus the short continuum integration time needed (i.e. $\sim$ 1 ms. The exact time will depend on how many wavelength bands we devide the continuum into). The optimum number of mirrors for forming a cophased sub-array will be determined through the combined modeling and experimental investigations during the design study of SI, of which FIT will serve as a first step.

An added advantage often cited for a Fizeau type of configuration is its wide instantaneous field of view, determined solely by the off-axis optical performance of the system and by the size of the detector. A Michelson type of interferometer potentially can also achieve wide field of view utilizing a large format detector array [3], though long strokes of the delay-lines are needed, and, in the case of small number of mirrors, numerous interferometer reconfigurations as well.

## 3. OBJECTIVES AND DESIGN OF THE FIT

The Fizeau Interferometer Testbed in its final form will include up to 30 distinct mirrors, which are fully articulated and will be commandable automatically by a closed-loop feedback system.

The main objectives of the FIT are the following:

- It is designed to explore the principles of and requirements for the SI mission concept, as well as other Fizeau type interferometers.
- It will utilize 7-30 separate apertures, each with 5 degrees of freedom (tip, tilt, piston, as well as 2d translations) in a sparse distribution.
- It has the chief goal of demonstrating closed-loop control of articulated mirrors and the overall system to keep beams in phase and optimize imaging.
- It will be used to determine the system requirements for accuracy and stability of the optics and metrology, and for vibration and stray light. We will then translate these into requirements for station keeping and formation-wide metrology for the SI.
- It will enable critical assessment of various image reconstruction algorithms, including phase retrieval [4] [5], phase diversity [6], CLEAN [7] [8], Maximum Entropy [9] [10], etc. for utility and accuracy by application to real data.
- It will also be used to investigate the optimal sampling methodology of the Fourier uv-plane, and the optimal implementation of that sampling via time-efficient and propellant-efficient reconfigurations of the array.
- It will be used to confirm achievable sensitivities for given Fourier uv-plane sampling and coverage, and determine the optimal number of collectors, dish size, and formation.

A schematic drawing of the FIT design, which allow the above objectives to be realized, is given in Figure 1. The first prototype of FIT will operate at the optical wavelengths and use a minimum-redundancy array [11] [12] for the primary mirror segments. An extended-scene film is illuminated by the light from the source assembly. The scene is located in the focal plane of the collimator mirror assembly, which consists of a hyperboloidal secondary and an off-axis paraboloidal primary. The collimated light is then intercepted by the elements of the spherical primary mirror array, which relay it to the oblate ellipsoid secondary mirror, which finally focuses it onto the image focal plane. An optical trombone arrangement near the focal plane allows an in-focus as well as an out-of-focus image to be simultaneously recorded on two CCD arrays for phase-diversity wavefront sensing analysis. An MS Windows box contains National Instrument devices for commanding piezo actuators which control the articulated primary mirror elements, and for controlling the data acquisition by the CCD arrays. The acquired data is relayed to a back-end parallel computer for analysis, and the result is fed back to the Windows box to be translated into the actuator control signals.

In the next few sections we describe in more details the FIT optics design, wavefront sensing approaches, as well as the control loop.

## 4. OPTICS DESIGN

The FIT optics design is chosen to incorporate many of the essential elements of the SI instrument. Specifically, the primary mirror of the imager assembly is chosen to be of spherical shape, which significantly reduces the manufacturing cost of the mirror segments, and also simplifies the external metrology system which will be used during the initial phasing of the array as well as for continuous monitoring of the optics stability. We now describe the procedures used to arrive at the FIT optics design.

We have already available to us an off-axis paraboloid collimator which has a 3 meter focal length, 12 inches of optical-quality aperture with a de-center distance of 200 mm. Therefore all the subsequent optics design assumes the use of this piece of existing optics. Through an optimization process of the collimator and imager optics design we found that the maximum usable aperture is limited (by aberration as well as by blockage) to about 10 inches.

*Object and Image Mapping*

Assume the film used for the object has grain size of 10 $\mu$m, and the CCD used for detection has pixel size also about 10 $\mu$m. For the object end, it is desired that the core of the point-spread-function (PSF) formed by the collimator optics is represented by 4 grains on the film. For a D=10 inch or 0.254 meter diameter aperture, and a median operating wavelength of 0.6$\mu$m, we have

$$2.44 \frac{\lambda}{D} \cdot F_1 = 40 \ \mu m, \quad (4)$$

which gives an the effective focal length of the collimator optics of $F_1 = 6.9$ meters, and an effective F-number about 27. Since the off-axis paraboloid we have has a focal length of 3 meters, a secondary mirror for the collimator assembly is needed to obtain the desired overall focal length. The object size we are eventually interested in imaging is about 2.4 mm, which is represented by 60 samples of the core of the collimator PDF, 120 samples of the effective linear resolution elements, and 240 samples of the film grain size, if we use the above collimator optics parameters.

On the image end, the collector mirror (which consists of the assembly of spherical mirror segments) is also chosen to have a diameter of D=0.254 meter. Choosing the receiving optics F-number to be 40 (dictated by the desired object-image magnification), and thus a focal length of $F_2 = 10.4$ meters, the core of the PSF in the image plane is

$$2.44 \frac{\lambda}{D} \cdot F_2 = 60 \ \mu m \quad (5)$$

Thus, there are about 6 CCD pixels within the central core of the image PSF.

The angular resolution of both the collimator optics and the collector optics is given by $1.22 \lambda/D = 2.9$ microradian. The magnification of the system as given by image-size/object-size is 1.5. The image formed on the CCD of the object of 2.4 mm in size is therefore 3.6 mm in linear size. A CCD of 4 mm x 8 mm in format will be able to contain both the in-focus and out-of-focus images during the phase-diversity imaging process.

*Analytic Design Equations*

The analytical design equations of the generic two-mirror systems are given in [13], which are based on those originally derived by [14].

For a two-mirror telescope (Figure 2), the equation for the primary mirror can be written as

$$x_1 = y_1^2/4f_1 + (1+b_1)y_1^4/8(2f_1)^3 + ..., \quad (6)$$

and for the secondary

$$x_2 = y_2^2/4f_2 + (1+b_2)y_2^4/8(2f_2)^3 + ..., \quad (7)$$

where $x_1$, $y_1$, $x_2$, $y_2$ measures the distance from the mirror vertex in the axial as well as vertical direction for the primary and secondary mirror, respectively, and $b_1$, $b_2$ measures the asphericity (they are the so-called "conic" constants in the ZEMAX raytracer. For a sphere, b = 0; for a paraboloid, b=-1; and for a conic with eccentricity $\epsilon$, $b = -\epsilon^2$); finally, $f_1, f_2$ denotes the focal length of the primary and secondary mirror, respectively.

The effective focal length of the combined two-mirror system is given by

$$1/f = 1/f_1 + 1/f_2 - d/f_1 f_2 \quad (8)$$

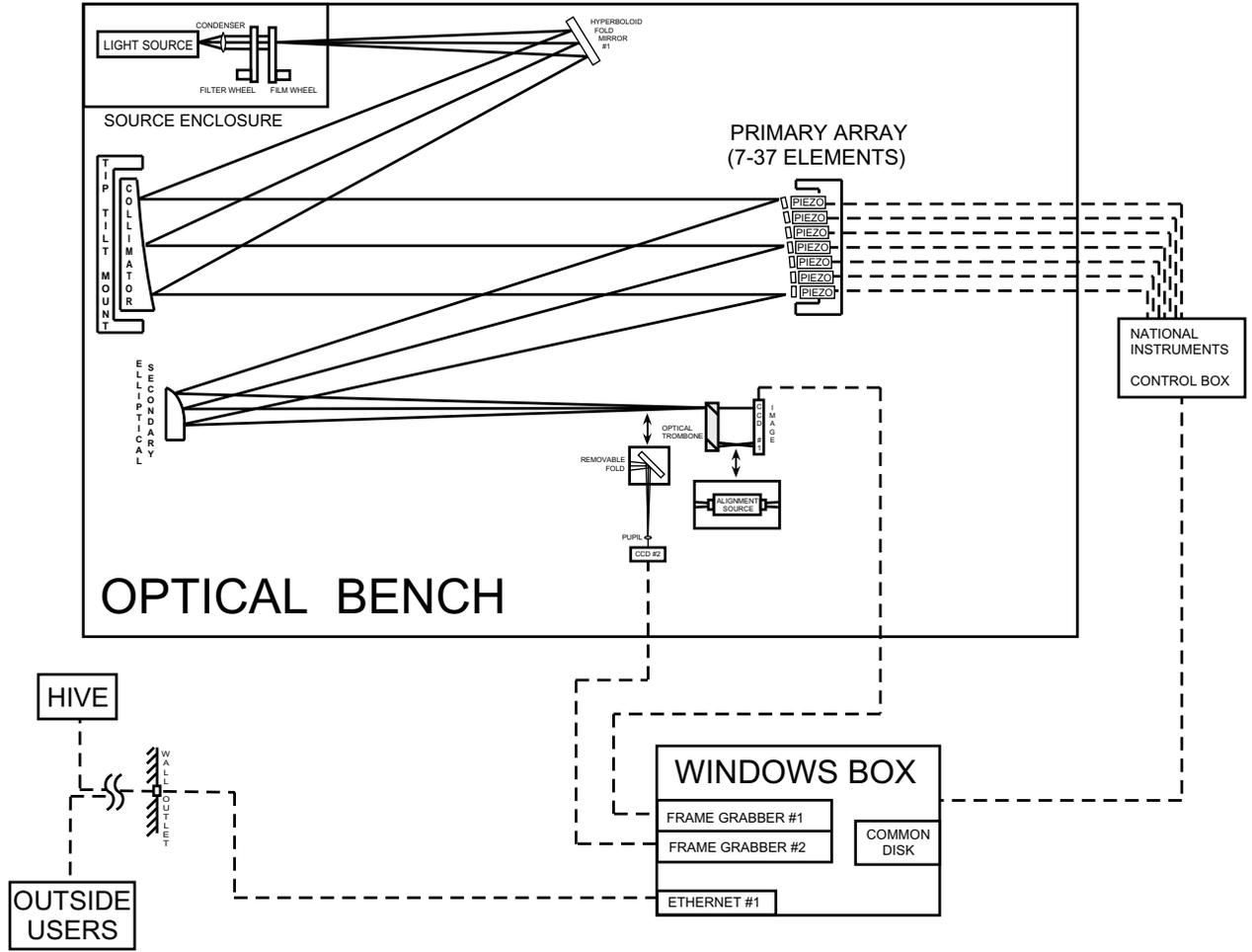

**Figure 1**. Fizeau Interferometer Testbed Schematic

where d is the separation between the two mirrors. Other useful relations are

$$(f + f_1)d = f_1(f - e), \qquad (9)$$

$$-(f - f_1)f_2 = f_1(d + e) = f(f_1 - d), \qquad (10)$$

where e is the back focal length.

Assume the entrance pupil coincides with the primary, the aberration coefficients of the two-mirror system are given by

$$B = \frac{1 + b_1}{8f_1^3} - \{b_2 + \left(\frac{f + f_1}{f - f_1}\right)^2\} \frac{(f - f_1)^3(f_1 + e)}{8f^3 f_1^3 (f + f_1)}, \qquad (11)$$

$$F = \frac{1}{4f^2} - \{b_2 + \left(\frac{f + f_1}{f - f_1}\right)^2\} \frac{(f - f_1)^3(f - e)}{8f^3 f_1^2 (f + f_1)}, \qquad (12)$$

$$C = \frac{f + e}{2f(f_1 + e)} - \qquad (13)$$

$$\{b_2 + \left(\frac{f + f_1}{f - f_1}\right)^2\} \frac{(f - f_1)^3(f - e)^2}{8f^3 f_1(f + f_1)(f_1 + e)},$$

and

$$C - D = \frac{1}{2f_1} + \frac{1}{2f_2}, \qquad (14)$$

where B is the coefficient of spherical aberration, F is the coefficient of tangential coma, where $\phi$ is the semi-angular field, and C is the coefficient of astigmatism.

For the FIT imager assembly, since the primary mirror is of spherical shape, we have $b_1 = 0$. The additional requirement of zero on-axis spherical aberration of the combined two-mirror system ($B = 0$ in equation 11) leads to $b_2$ of

$$b_2 = \frac{f_1 + f}{f_1 + e}\left(\frac{m}{m - 1}\right)^3 - \left(\frac{m + 1}{m - 1}\right)^2 \qquad (15)$$

where $m \equiv f/f_1$ is the magnification of the two-mirror system.

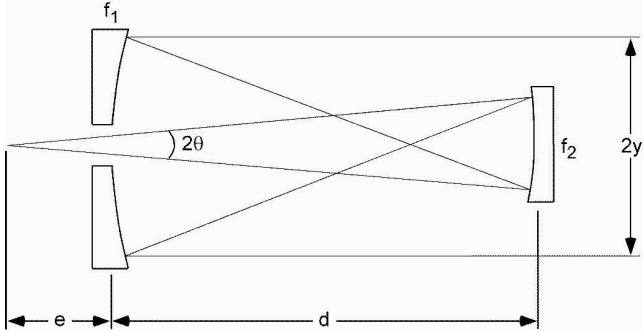

**Figure 2**. Two-Mirror Telescope

The sign convention for the various quantities in the above equations are: the focal length for the individual mirrors are always greater than zero. For the overall system, if it is a Cassegrain type (concave/convec), then $f > 0, m > 0$, and for a Gregorian type (concave/concave), $f < 0, m < 0$.

Since $b = -\epsilon^2$, we have that the various shapes of the surface of revolution are given by

- $\epsilon < -1, b < -1$, for hyperboloid.
- $\epsilon = -1, b = -1$, for paraboloid.
- $< -1 < \epsilon < 0, -1 < b < 0$, for prolate ellipsoid.
- $\epsilon = 0, b = 0$, for sphere.
- $\epsilon$ imaginary, $b > 0$, for oblate ellipsoid.

The above analytical design equations have been used to arrive at the initial optics design, which is further optimized using the commercial ray-tracer ZEMAX.

*ZEMAX-Optimized Optical Configuration*

Figure 3 shows the overall optics layout for the collimator and the imager assemblies. We have shown here the full mirror sizes out of which the off-axis pieces are cut to form the actual optical elements.

The specific parameters of the optics are:

- *Object*: diameter 2.4mm, distance to next element 1.813m.
- *Collimator secondary*: radius of curvature 2.863m, conic -6.65, diameter 84mm, decenter 53.5mm, distance to the next element 2.2m.
- *Collimator primary*: radius of curvature 6m, conic -1, (usable) diameter 254mm, decenter 200mm, distance to the next element 2.5m.
- *Imager primary*: radius of curvature 4m, (usable) diameter 254mm, decenter 280, distance to the next element 1.414m.
- *Imager secondary*: radius of curvature 1.463m, conic 4.5, diameter 110mm, decenter 81mm, distance to the next element 2.948m.
- *image*: diameter 3.7mm.

Raytrace shows that the performance of the overall optics from the object plane to the image plane is diffraction-limited over most of the 2.4mm x 2.4mm (or 1.2' x 1.2') field-of-view.

## 5. RADIOMETRY

The purpose of the radiometry calculation is to verify that for a particular set of light source, system optics and camera combination, the sensitivity of the detection system matches that of the received light level after passing through the optics system. Specifically, the resulting illuminance or irradiance at the location of the CCD should be sufficient to power the full-well capacity of the CCD given its photopic response.

It is common to use auxiliary optics after a light source to vary the distribution of the light so as to achieve optimum illumination of the object. Suppose that the illumination optics (condenser lens, or microscope objective) has mapped the source illumination spot of area $A_{source}$ into the exact size of a pinhole $A_{pinhole}$, with $A_{pinhole}/A_{source} = M^2$, where M is the linear magnification of the illumination optics. Due to the conservation of the etendue, $A_{source} \cdot \Omega_{source} = A_{pinhole} \cdot \Omega_{pinhole}$. Therefore

$$\Omega_{pinhole} = \frac{1}{M^2}\Omega_{source}. \quad (16)$$

Denote the pickup mirror (one of the mirror segments on the primary) angular size as $\Omega_{pickup}$, then the power it receives is related to the power transmitted by the pinhole by

$$\text{Power}_{pickup} = \text{Power}_{pinhole} \cdot \frac{\Omega_{pickup}}{\Omega_{pinhole}}. \quad (17)$$

Now

$$\text{Power}_{pinhole} = A_{pinhole} \cdot E_{pinhole} = A_{pinhole} \cdot \frac{E_{source}}{M^2}, \quad (18)$$

where $E_{pinhole}$ and $E_{source}$ denote power density in the unit of $Wm^{-2}$.

Therefore

$$\text{Power}_{pickup} = A_{pinhole} \cdot \frac{E_{source}}{M^2} \cdot \frac{\Omega_{pickup}}{\Omega_{source}/M^2} \quad (19)$$

$$= A_{pinhole} \cdot E_{source} \cdot \frac{\Omega_{pickup}}{\Omega_{source}},$$

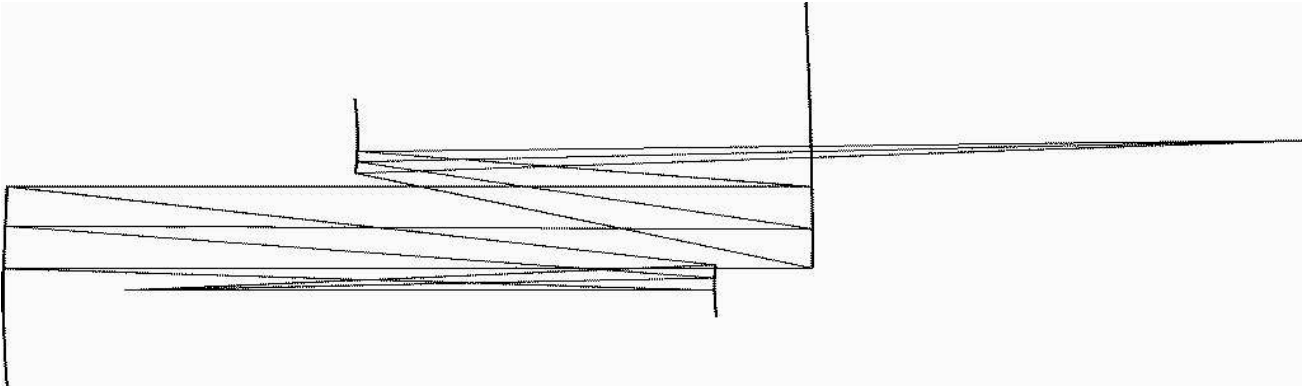

**Figure 3**. Layout of the overall optical system showing the full mirrors out of which the off-axis pieces will be cut.

which is independent of M! We have thus shown that the magnification of the illumination optics has no effect on the power received by the pickup mirror, as long as $\Omega_{pinhole}$ encompasses all of the pickup mirrors. We also see from the result of (5) that we have three ways to increase the pickup power: (1) increase the pickup mirror size $\Omega_{pickup}$, (2) increase the source surface brightness $E_{source}/\Omega_{source}(Wm^{-2}sr^{-1})$, (3) increase the pinhole size $A_{pinhole}$.

In our case, the equivalence of $A_{pinhole}$ is the area of the object film (or the sub-areas of it corresponding to the resolution elements), if we choose a condenser system to focus the light of the source onto an area roughly the size of the film, and check to make sure that the solid angle of the output beam encompasses the extent of the primary mirror. Our choice of the resolution and magnification already fixed the $A_{pinhole}$ value. The sparcity of the Fizeau array on the other hand dictates the size of the pick-up mirrors. Moreover, we see that it is the source surface brightness $E/\Omega$ (itself proportional to the fourth power of the color temperature of the lamp filament), that enters into the received power relation. Increasing the lamp output by increasing the wattage, which leads to an increase in the surface area of the filament, does not have a direct bearing on the received power by the pickup mirror. Therefore, light-starving applications such as FIT require the use of sources with higher color temperature, i.e. Xenon light source rather than halogen [15].

## 6. WAVEFRONT SENSING AND CONTROL

FIT will be required to maintain high quality imaging with multiple array elements. The misalignments and deformations of the optics and mounts will cause optical phase aberrations, i.e. deviation of the optical wavefront from the ideal optical wavefront (wavefront error). In order to operate in closed loop we must first sense the wavefront error (wavefront sensing) and then correct it (optical control).

A multitude of methods are available for wavefront sensing. Some examples of these methods are: (i) interferometry, (ii) Shack-Hartmann sensing, and (iii) phase retrieval/diversity. Interferometry requires additional optics and detectors, thereby increasing system complexity and cost. Furthermore, this configuration does not "see" the entire optical train of the science instrument, thus the error sensed is incomplete. Moreover, a fraction of the photons are picked off prior to the science focal plane to be used for the wavefront sensing, thus lowering the number of useful science photons if the wavefront sensing and the science observations are done in the same wavelength. The Shack-Hartmann approach uses an array of lenslets and senses the slope of the wavefront and hence is not sensitive to relative piston errors between the array elements. Phase retrieval [4] [5] and phase diversity [6] are image-based methods which utilize the science image(s) along with an optical model of the system to determine the phase. It does not require hardware other than what is in the system. Phase retrieval was successfully used on the Hubble Space Telescope [16], [17] and is proposed as the baseline wavefront sensing method for the James Webb Space Telescope [18] [19].

Despite the success of the application of phase-retrieval and phase-diversity to the wavefront sensing and correction of deformable as well as segmented mirrors, their applicability to the sparse aperture array with a low fill factor is yet to be established. An external metrology system will also be set up to monitor the stability of the optics.

Once the wavefront is sensed it is subsequently "fit" to the response matrix of the actuators. The response matrix consists of the eigenmodes of the correctable degrees of freedom of the system[20] The control loop will initially operate around 1 Hz, and will be gradually improved as the system performance and the wavefront retrieval calculations are optimized.

## 7. ACKNOWLEDGMENT

We thank our collaborators at NASA's GSFC: Alan Centa, Bill Danchi, Lisa Mazzuca, and Susan Neff; at Sigma Space Corporation, Paul Cottle; at Mink Hollow Systems, Dave McAndrew; at the Naval Research Laboratory, Tom Armstrong, Dave Mozurkewich and Tom Pauls; at the University of Maryland, Lee Mundy; at LMATC, Carolus Schrijver; at the Space Telescope Science Institute, Jay Rajagopal and Ron Allen; and at the CfA, Margarita Karovska, for their contributions to the development of the testbed and the Stellar Imager mission concept.

This research is supported in part by NASA ROSS/SARA funding and by NASA Goddard Space Flight Center's internal research and development fund. The Naval Research Laboratory's participation in the FIT and SI development is supported in part by the Office of Naval Research.

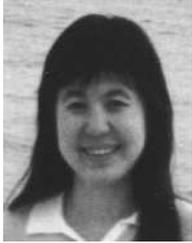
**Xiaolei Zhang** is an Astrophysicist at the US Naval Research Laboratory. Her research interests span the area of galaxy evolution and cosmology, optical interferometry, as well as the experimental studies of laser cooling and matter wave sensors. Dr. Zhang obtained MS and PhD degrees in electrical engineering from the University of California at Berkeley in 1987 and 1992, respectively. Since graduation she has worked at the Harvard-Smithsonian Center for Astrophysics first as a postdoc and subsequently as a staff astronomer, as well as at NASA's Goddard Space Flight Center as an employee of Raytheon ITSS and SSAI. She joined the Naval Research Laboratory in July 2002. Dr. Zhang is a member of the American Astronomical Society.

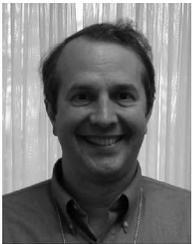
**Kenneth Carpenter** is currently the Project Scientist for Hubble Space Telescope Operations and splits his time between those duties, leading the Stellar Imager mission concept study, and scientific research. Dr. Carpenter's scientific interests include studies of the chromospheres, transition regions, winds and circumstellar shells of cool stars, as well as the calculation of model atmospheres and synthetic spectra and investigations of line fluorescence processes; hardware interests include development and operations of UV spectroscopic instruments and large baseline space interferometers.

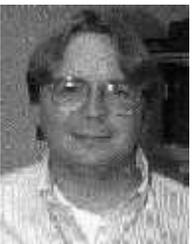
**Richard Lyon** is an optical scientist at NASA's Goddard Space Flight Center with approximately 15 years experience in the areas of computational optics, wavefront sensing, optical control systems, intereferometric imaging, coronagraphy and image enhancement. He holds 10 achievement awards for NASA, NOAA and DOD flight missions. Mr. Lyon holds a BS in Physics from the University of Massachusetts, an MS in Optics from the University of Rochester and has completed work towards the Ph.D degree in Optics while at the University of Rochester. He is a member of the Optical Society of American and the Society for Photo-Instrumentation Engineers.

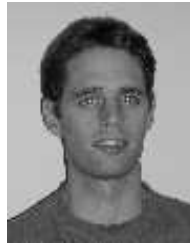
**Hubert Huet** is an optical engineer working for the Sigma Space Corporation. He is currently involved in the design of various optical systems, among them a new kind of sun sensor using holographic techniques. H. Huet obtained a MS degree in optical engineering from the Eco le Superieure d'Optique in France.

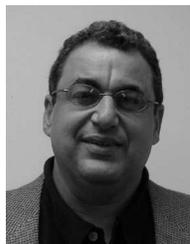
**Joe Marzouk** is the Vice President and director of Optical systems of Sigma Space Corporation. He has supported a number of programs and is the Lead optical engineer on the Earth Observing System (EOS) GLAS program, co-investigator on the Holographic Airborne Rotating Lidar Imaging Experiment (HARLIE) program, optical engineer on the Mars Global Surveyor (MGS) Mars Observer Laser Altimeter (MOLA) program, member of the Hubble Space Telescope Independent Verification Team, lead opto-mechanical engineer on PAMS, and alignment consultant on AXAF-S and Astro-E programs. He serves as Principal Investigator for the Advanced holographic Sun sensor and holds a number of technical patents.

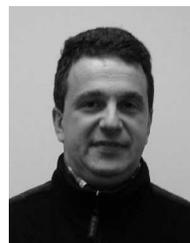
**Gregory Solyar** received a BS degree in radio metrology from Odessa Measuring College of Standards Committee of USSR in 1981, a MS degree in electrical engineering from Odessa State Polytechnic Institute in 1986, and a MS degree in electrical and computer engineering from Johns Hopkins University in 1998 where his major interest was in nonlinear optics and quantum electronics. He is currently working toward a Ph.D. degree in the Department of Computer Science and Electrical Engineering of UMBC where the research interest is in imaging methods and algorithms. In December 2000 he joined NASA CESDIS and became an associate research scientist for NASA GEST in June of 2000. He has over 15 years of working experience in industry and research institutions.